\documentclass[%
 reprint,
 amsmath,amssymb,
 aps,prl
]{revtex4-2}
\usepackage{graphicx}
\usepackage{dcolumn}
\usepackage{bm}
\usepackage{hyperref}
\usepackage{color}

\hypersetup{
    colorlinks=true,
    linkcolor=blue,
    filecolor=magenta,      
    urlcolor=magenta,
    pdftitle={Paper},
    pdfpagemode=FullScreen,
    }

\usepackage{soul}

\begin{document}

\preprint{APS/123-QED}



\title{A Schmidt decomposition approach to quantum thermodynamics}

\author{André Malavazi}
 \email{andrehamalavazi@gmail.com}
\author{Frederico Brito}%
 \email{fbb@ifsc.usp.br}
\affiliation{%
 Instituto de Física de São Carlos, Universidade de São Paulo, C.P. 369, 13560-970 São Carlos, SP, Brasil
}%

\date{\today}



\begin{abstract}

The development of a self-consistent thermodynamic theory of quantum systems is of fundamental importance for modern physics. Still, despite its essential role in quantum science and technology, there is no unifying formalism for characterizing the thermodynamics within general autonomous quantum systems, and many fundamental open questions remain unanswered. Along these lines, most current efforts and approaches restrict the analysis to particular scenarios of approximative descriptions and semi-classical regimes. Here we propose a novel approach to describe the thermodynamics of arbitrary bipartite autonomous quantum systems based on the well-known Schmidt decomposition. This formalism provides a simple, exact and symmetrical framework for expressing the energetics between interacting systems, including scenarios beyond the standard description regimes, such as strong coupling. We show that this procedure allows a straightforward identification of local effective operators suitable for characterizing the physical local internal energies. We also demonstrate that these quantities naturally satisfy the usual thermodynamic notion of energy additivity.

\end{abstract}



\maketitle




During the last decades, there have been many efforts into the extension of thermodynamics both to encompass microscopic systems out of equilibrium \cite{jarzynski1997nonequilibrium,seifert2012stochastic} and to include genuine quantum phenomena. Along with the state-of-the-art capability of controlling fragile quantum systems in a wide variety of physical platforms, this context has paved the way for the current strategic endeavours to develop a self-consistent thermodynamic theory of quantum systems \cite{gemmer2009quantum,vinjanampathy2016quantum,binder2018thermodynamics,deffner2019quantum}. Interestingly, despite all recent extensive research and advances in quantum thermodynamics \cite{batalhao2014experimental,an2015experimental,rossnagel2016single,klatzow2019experimental}, some open questions concerning central aspects of the theory are exceptionally notorious. In particular, there are no universally accepted general definitions for the quantum counterparts of the most basic thermodynamic quantities and, therefore, no formalism consolidated for a quantum thermodynamic description suitable for arbitrary quantum systems. This context poses a fundamental restriction to many cases of interest and highlights a core issue of the field.

While there is no ambiguity in defining proper thermodynamic quantities for macroscopic systems in equilibrium, fundamental subtleties arise once one must account for non-negligible interactions and intrinsically quantum properties of systems out-of-equilibrium. In this sense, the identification of physical local internal energy and quantum thermodynamic entropy is not as straightforward as in the classical scenario since it is unclear how to proceed in the presence of these notions. Although different definition approaches have been reported in the literature \cite{Alicki1979,Weimer2008,Alipour2016a,Valente2018,Colla2021a,Esposito_2010,Polkovnikov2011,Entropy2019}, this task remains elusive and far from consensus. Naturally, any attempt to describe energy exchanges, such as work and heat, inherits this fundamental uncertainty and brings its own conceptual issues and challenges \cite{Talkner2007,Elouard2016b,Perarnau-Llobet2017,Ahmadi2019a,Alipour2019b,Bernardo2020}.


Additionally, most current approaches in quantum thermodynamics commonly rely on a particular class of assumptions and approximations, which inevitably limits their application to a small set of regimes. In particular, it is often assumed weakly-coupled systems, Markovian dynamics, and semi-classical external control \cite{Alicki1979,kosloff2013quantum,kosloff2019quantum,dann2021open}. Despite being well-suitable for specific scenarios, such assumptions clearly exclude several relevant situations.

Along these lines, it is clear that the current status of the theory is restrictive and does not allow the description and characterization of thermodynamic processes for general contexts, especially for autonomous quantum systems and in the presence of strong coupling. In this respect, many efforts have been attempting to expand quantum thermodynamics to these broader conditions \cite{Weimer2008,Hossein-Nejad2015,Alipour2016a,dann2021quantum,esposito2015nature,PhysRevB.98.134306,perarnau2018strong,strasberg2019repeated,Rivas2020,Colla2021a}. Nevertheless, there is no general framework suitable for arbitrary quantum systems.


This work is inserted in this context. Here, we introduce a novel approach for analyzing the energetic exchange occurring between parts of a general autonomous quantum system. This proposal is exact and based on the well-known mathematical procedure of the Schmidt decomposition for bipartite systems. Interestingly, despite being simple and providing a powerful statement, it is still not explicitly explored in the context of quantum thermodynamics. This framework will allow us to describe the dynamics and energetics within generic interacting subsystems in a symmetrical fashion, i.e., regardless of their individual properties, details, and dimension, they will be treated on equal footing. In addition, it will not require any complementary hypotheses and approximations, such as the commonly used ones concerning the Hamiltonian structure, interaction regimes, and type of dynamics, i.e., strict energy conservation, weak-coupling, and Markovianity. Formally, we will introduce time-dependent local effective Hamiltonians that naturally embrace both their respective bare ones and the contributions of the interaction term. These elements will be identified as the representative operators for characterizing the subsystem's physical internal energies and will allow us to extend the usual classical thermodynamic notion of energy additivity to general interacting quantum systems. Such identification represents one of our main results.



\section{Results}

\paragraph{System description}

As a quantum universe, we consider a finite and isolated quantum system composed of two interacting subsystems with dimensions $d^{(1)}$ and $d^{(2)}$. In addition, let us assume that - without any loss of generality - $d^{(1)}\leq d^{(2)}$. The whole system Hamiltonian is given by 
\begin{equation}
\hat{H}^{(0)}:=\hat{H}^{(1)}\otimes\hat{1}^{(2)}+\hat{1}^{(1)}\otimes\hat{H}^{(2)}+\hat{H}_{int},\label{Total Hamiltonian}
\end{equation}
where $\hat{H}^{(k)}$ is the local bare Hamiltonian of subsystem $(k)$ with $k=1,2$ and
$\hat{H}_{int}$ is the term encompassing all interactions between them. 

At any time $t$, the universe's quantum state is depicted
by a pure state represented by the state vector $|\Psi(t)\rangle$, whose dynamics is unitary and governed by the usual Schrödinger equation $i\hbar\frac{d}{dt}|\Psi(t)\rangle=\hat{H}^{(0)}|\Psi(t)\rangle$. The local state description for each subsystem is expressed by the reduced density matrix of the universe, such that $\hat{\rho}^{(1,2)}(t)\equiv tr_{2,1}\{|\Psi(t)\rangle\langle\Psi(t)|\}$. No additional assumptions regarding the universe, its constituents or the Hamiltonians will be considered.



\paragraph{Schmidt decomposition approach}

Naturally, $|\Psi(t)\rangle$ can be written according to any conceivable basis. Nevertheless, the well known Schmidt decomposition \cite{nielsen10} provides the following particular and convenient form 
\begin{equation}
    |\Psi(t)\rangle=\sum_{j=1}^{d^{(1)}}\lambda_{j}(t)|\varphi_{j}^{(1)}(t)\rangle\otimes|\varphi_{j}^{(2)}(t)\rangle,\label{SchmidtDecomposition}
\end{equation}
for every instant $t$, where $\{\lambda_{j}(t)\geq0;j=1,...,d^{(1)}\}$ and $\{|\varphi_{j}^{(k)}(t)\rangle;j=1,...,d^{(1)}\}$ are the time-local Schmidt coefficients and local Schmidt basis of subsystem $(k)$, respectively. The representation above is compelling and useful for a number of reasons: notice that despite a potential huge discrepancy between $d^{(1)}$ and $d^{(2)}$ there is a single sum bounded by the smallest dimension in question, by hypothesis $d^{(1)}$; apart from that, there is a one-to-one relation between the Schmidt basis elements, i.e., for each Schmidt ket $|\varphi_{j}^{(1)}(t)\rangle$ from subsystem $(1)$ there is a related ket $|\varphi_{j}^{(2)}(t)\rangle$ from $(2)$; also, it is guaranteed that the Schmidt coefficients are unambiguously defined, while the Schmidt basis are unique up to eventual degenerate coefficients and a phase degree of freedom, in the sense that equation (\ref{SchmidtDecomposition}) is invariant over simultaneous local phases changes; additionally, it gives all necessary information for representing the subsystem's local density matrices $\hat{\rho}^{(k)}(t)$, i.e., their simultaneous spectral decomposition is given by the Schmidt coefficients squared $\{\lambda_{j}^{2}(t)\}_{j}$ and the Schmidt basis $\{|\varphi_{j}^{(k)}(t)\rangle\}_{j}$, such that $\hat{\rho}^{(k)}(t)=\sum_{j=1}^{d^{(1)}}\lambda_{j}^{2}(t)|\varphi_{j}^{(k)}(t)\rangle\langle\varphi_{j}^{(k)}(t)|$. In particular, notice that both local state representations are - in general - mixed state and have the same spectrum whenever the universe is pure.

The autonomous time evolution executed by the whole system at any time interval $[t_{0},t_{1}]$ also induces the local dynamic changes, which, in turn, are mainly non-unitary and strongly influenced by the presence of the interaction term $\hat{H}_{int}$. Let us now focus on the changes in the Schmidt decomposition elements and define the local dynamical maps $\tilde{\mathcal{U}}^{(k)}:\mathcal{H}^{(k)}\rightarrow\mathcal{H}^{(k)}$ as the operators characterizing the Schmidt basis dynamics, i.e., $|\varphi_{j}^{(k)}(t)\rangle=\tilde{\mathcal{U}}^{(k)}(t,t_{0})|\varphi_{j}^{(k)}(t_{0})\rangle$, for any $t\geq t_{0}$, with $\underset{t\rightarrow t_{0}}{\lim}\,\tilde{\mathcal{U}}^{(k)}(t,t_{0})=\hat{1}^{(k)}$. Since the Schmidt basis at distinct times corresponds to different orthonormal basis for the same Hilbert space, the relationship above is unitary, and one can define the local time-translation generators $\tilde{H}^{(k)}(t)=\tilde{H}^{(k)\dagger}(t)$ of the Schmidt basis $\{|\varphi_{j}^{(k)}(t)\rangle\}_{j}$, such that $i\hbar\frac{d}{dt}\tilde{\mathcal{U}}^{(k)}(t,t_{0})=\tilde{H}^{(k)}(t)\tilde{\mathcal{U}}^{(k)}(t,t_{0})$,
and, therefore,
\begin{equation}
    i\hbar\frac{d}{dt}|\varphi_{j}^{(k)}(t)\rangle=\tilde{H}^{(k)}(t)|\varphi_{j}^{(k)}(t)\rangle,\label{Schmidt basis dynamics}
\end{equation}
for all $j$. Alternatively, $\tilde{H}^{(k)}(t)$ can be simply cast as
\begin{equation}
    \tilde{H}^{(k)}(t)\equiv i\hbar\sum_{j=1}^{d^{(k)}}\left( \frac{d}{dt}|\varphi_{j}^{(k)}(t)\rangle\right)\langle\varphi_{j}^{(k)}(t)|.\label{LocalEffectiveHamiltonian}
\end{equation}
Still, we can go one step further and show that equation (\ref{LocalEffectiveHamiltonian}) above can be directly related to the local bare Hamiltonian $\hat{H}^{(k)}$ as follows (see Methods section)
\begin{equation}
    \tilde{H}^{(k)}(t)=\hat{H}^{(k)}+\hat{H}_{LS}^{(k)}(t)+\hat{H}_{X}^{(k)}(t),\label{LocalEffectiveHamiltonian2}
\end{equation}
i.e., $\tilde{H}^{(k)}(t)$ can be split into the sum of three distinct elements, including the bare Hamiltonian. The additional operators are responsible for the time-dependency of $\tilde{H}^{(k)}(t)$, where $\hat{H}_{LS}^{(k)}(t)$ is a general Lamb-shift like term, in the sense that it is diagonal in the bare eigenbasis, i.e., $[\hat{H}_{LS}^{(k)}(t),\hat{H}^{(k)}]=0$ for all $t$, and $\hat{H}_{X}^{(k)}(t)$ contains only non-diagonal elements. Note that in the absence of interactions between the partitions, both subsystems would individually evolve in time in a unitary fashion. Under these circumstances, the role of the time-translation generator of the Schmidt basis would be naturally assigned to the respective local bare Hamiltonians. That is, the free evolution of the Schmidt basis would be simply $|\bar{\varphi}_{j}^{(k)}(t)\rangle=e^{-\frac{i}{\hbar}\hat{H}^{(k)}(t-t_{0})}|\varphi_{j}^{(k)}(t_{0})\rangle$. In this sense, it is clear that \textbf{(i)} the elements $\hat{H}_{LS}^{(k)}(t)$ and $\hat{H}_{X}^{(k)}(t)$ are direct local byproducts of the interaction term and vanish for non-interacting systems; \textbf{(ii)} equation (\ref{LocalEffectiveHamiltonian2}) acts as an effective Hamiltonian. Thus, from now on, $\tilde{H}^{(k)}(t)$ will be referred to as the \textit{local effective Hamiltonian} of subsystem $(k)$.

Also, one finds that the local effective Hamiltonians are responsible for guiding the unitary component of the dynamics of $\hat{\rho}^{(k)}(t)$, i.e.,
\begin{equation}
    i\hbar\frac{d}{dt}\hat{\rho}^{(k)}(t)=[\tilde{H}^{(k)}(t),\hat{\rho}^{(k)}(t)]+\mathfrak{\hat{D}}_{t}^{(k)}\hat{\rho}^{(k)}(t),\label{Master Equation}
\end{equation}
where $\mathfrak{\hat{D}}_{t}^{(k)}\hat{\rho}^{(k)}(t)=i\hbar\sum_{j=1}^{d^{(1)}}\left(\frac{d}{dt}\lambda_{j}^{2}(t)\right)|\varphi_{j}^{(k)}(t)\rangle\langle\varphi_{j}^{(k)}(t)|$ is the non-unitary part that explicitly depends on the Schmidt coefficients time dependency.



\paragraph{Internal energies and additivity}

By construction, in the Schrödinger picture, the time-translation generator of isolated and closed quantum systems \textit{is} the functional of energy. Thus, the total internal energy, $U^{(0)}$, contained within the universe under consideration is completely specified by the bare Hamiltonian $\hat{H}^{(0)}$, i.e., one can readily identify
\begin{equation*}
    U^{(0)}\equiv\langle\hat{H}^{(0)}\rangle
\end{equation*}
where $\langle.\rangle\equiv\langle\Psi(t)|(.)|\Psi(t)\rangle$. Additionally, since the bipartition is isolated, no energy flows inward or outward of the system. Along these lines, it is easy to see that the unitary evolution of $|\Psi(t)\rangle$ naturally guarantees internal energy conservation, such that
\begin{equation}
    \frac{d}{dt}U^{(0)}=0.
\end{equation}
Interestingly, it is unclear how to obtain a consistent and meaningful analogous for the interacting subsystems. In general, the notion of local internal energy is blurred by non-negligible interactions within the whole system, i.e., the total internal energy can be written as the sum of the individual local contributions of the expectation values of the bare Hamiltonians and the interaction term,
\begin{equation*}
    U^{(0)}=\langle\hat{H}^{(1)}\rangle(t)+\langle\hat{H}^{(2)}\rangle(t)+\langle\hat{H}_{int}\rangle(t).\label{total internal energy 1}
\end{equation*}
Thus, unless considering particular scenarios, it is clear that the interaction is indispensable for the computation of the universe's energy. Yet, its role is uncertain when one attempts to establish local observable for characterizing the subsystem's internal energy. Along these lines, given that the interaction term actively influences their local dynamics, it is reasonable to expect that its contribution should be shared between them, in one way or another.

In this context, it is desirable two relevant properties for any suitable definition of local internal energies: \textbf{(i)} be obtained by local measurements, i.e., associated with the expectation value of a local operator; \textbf{(ii)} be an additive quantity. While the first condition guarantees a local description and accessibility, the second also allows the intuitive picture of energy flowing from one system to another without including energetic sinks or sources, i.e., the sum of the local internal energies is a conserved quantity. These features, of course, are not trivial, especially because the interaction term acts on the whole Hilbert space, which means that it is a global property per se. This fact, however, suggests that an effective approach for describing local internal energy provides the most promising route. Otherwise, a global picture would be necessary for characterizing the energetic flux, which is impractical for most realistic scenarios.

We first remark that the local effective Hamiltonians presented earlier satisfy both desired features for being interpreted as the representative operators for characterizing the subsystem's physical internal energies. Thus, by identifying 
\begin{equation}
   U^{(k)}(t):=\langle\tilde{H}^{(k)}(t)\rangle,\label{local internal energy}
\end{equation}
the local internal energies become locally accessible properties, and one can show that their sum is equal to the constant universe's internal energy, such that 
\begin{equation}
   U^{(0)}=U^{(1)}(t)+U^{(2)}(t)\label{energy balance}
\end{equation}
and all energy flowing from subsystem $(1)$ is fully transferred to subsystem $(2)$, i.e., no additional energetic source is required (see Methods section).

Note that such description is general, symmetrical and does not rely on approximation, particular regimes and additional restrictive hypotheses concerning the subsystem dynamics, the Hamiltonian structure and the coupling strength. This identification represents one of our main results.



\paragraph{Phase gauge}

As mentioned earlier, the Schmidt basis is unique up to eventual degenerate Schmidt coefficients and a phase choice freedom. While the former is unimportant to our current purposes, the latter brings non-trivial consequences to our formalism. From a global perspective, it is easy to see that the simultaneous addition of local phases $\{\theta_{j}(t)\}_{j}\in\mathbb{R}$, such that \begin{equation} 
|\varphi_{j}^{\prime(k)}(t)\rangle = e^{(-1)^{k-1}i\theta_{j}(t)}|\varphi_{j}^{(k)}(t)\rangle,\label{gauge change}
\end{equation} 
maintains the universe state form unchanged, i.e., the phases cancel out and $|\Psi^{\prime}(t)\rangle=|\Psi(t)\rangle$. Such phase invariance expresses an internal gauge freedom choice inbuilt in the Schmidt decomposition that directly affects the identification of the local effective Hamiltonians, i.e., instead of describing the state vector with the basis $\{|\varphi_{j}^{(k)}(t)\rangle\}_{j}$, one can use $\{|\varphi_{j}^{\prime(k)}(t)\rangle\}_{j}$ and find its time-translation generator $\tilde{H}^{\prime(k)}(t)$. Given equation (\ref{LocalEffectiveHamiltonian}), the gauge transformation defined in equation (\ref{gauge change}) above implies the following expression for $\tilde{H}^{\prime(k)}(t)$ in terms of $\tilde{H}^{(k)}(t)$ and $\{\theta_{j}(t)\}_{j}$
\begin{equation}
    \tilde{H}^{\prime(k)}(t)=\tilde{H}^{(k)}(t)+(-1)^k\hbar\sum_{j=1}^{d^{(k)}}\left(\frac{d\theta_{j}(t)}{dt}\right)|\varphi_{j}^{(k)}(t)\rangle\langle\varphi_{j}^{(k)}(t)|.\label{Local effective Hamiltonian gauge change}
\end{equation}
Naturally, its expectation value also transforms accordingly, such that
\begin{equation}
    \langle\tilde{H}^{\prime(k)}(t)\rangle=\langle\tilde{H}^{(k)}(t)\rangle+(-1)^k\hbar\sum_{j=1}^{d^{(k)}}\lambda_{j}^{2}(t)\left(\frac{d\theta_{j}(t)}{dt}\right).\label{Expectation values Local effective Hamiltonian gauge change}
\end{equation}
Thus, it is clear that from a local point of view these phases are relevant and should be carefully scrutinized.

Note that a gauge change adds an extra diagonal term on the Schmidt basis that only depends on the time derivative of the phases. In general, these additional quantities change the local effective Hamiltonian structure in a way that both their eigenbasis and eigenvalues are affected, which also implies that the spectral gaps are not necessarily invariant. Still, as expected, such transformations maintain the description of the local density matrices and their dynamical equation (\ref{Master Equation}) unchanged, given that $[|\varphi_{j}^{(k)}(t)\rangle\langle\varphi_{j}^{(k)}(t)|,\hat{\rho}^{(k)}(t)]=0$ for all $j$ and $t$. More importantly, these terms shift the expectation values $\langle\tilde{H}^{(k)}(t)\rangle$ in such a way the change obtained by subsystem $(1)$ is compensated by the one acquired by subsystem $(2)$. Hence, both the universe's internal energy $U^{(0)}$ and the additivity property are guaranteed to be invariant over these modifications, i.e., $U^{\prime(0)}=U^{(0)}$ and $\langle\tilde{H}^{\prime(1)}(t)\rangle+\langle\tilde{H}^{\prime(2)}(t)\rangle=\langle\tilde{H}^{(1)}(t)\rangle+\langle\tilde{H}^{(2)}(t)\rangle$.

Hence, given the freedom of choosing the Schmidt basis for writing the state vector, it is clear that for the same bipartite quantum system, several local effective Hamiltonians are suitable for consistently describing the subsystem's dynamics, any observable physical quantity and the universe energetics. However, despite $\tilde{H}^{(k)}(t)$ possessing the desired properties for successfully fulfilling the role of physical local internal energies, such freedom also brings a significant ambiguity in identifying a single pair of preferential operators for quantifying them.


Notably, such mathematical freedom of adding local phases is not necessarily physically consistent for characterizing energy in general when seeking its association with the time-translation generator of the dynamics \footnote{In a private communication, Prof. Shang-Yung Wang presented an approach based on the introduction of gauge potentials such that the local effective Hamiltonian is gauge invariant. Even though this feature is quite neat, it does not preserve the association of the Hamiltonian as the time-translation generator, differing from the approach adopted here.}. As mentioned earlier, if there is no interaction between the bipartitions, they would behave independently as isolated quantum systems. In this scenario, the Schmidt basis dynamics would be described by $i\hbar\frac{d}{dt}|\bar{\varphi}_{j}^{(k)}(t)\rangle=\hat{H}^{(k)}|\bar{\varphi}_{j}^{(k)}(t)\rangle$, and the individual local internal energies would be readily identified as $U^{(k)}\equiv\langle\hat{H}^{(k)}\rangle$, where $\frac{d}{dt}U^{(k)}=0$. Along these lines, once considering the interaction term negligible, similar conclusions should be obtained regardless of the chosen gauge. In this respect, if $\{|\varphi_{j}^{(k)}(t)\rangle\}_{j}$ is  indeed the Schmidt basis such that $|\varphi_{j}^{(k)}(t)\rangle\rightarrow|\bar{\varphi}_{j}^{(k)}(t)\rangle$ when $\hat{H}_{int}\rightarrow0$, then it is guaranteed that $\tilde{H}^{(k)}(t)\rightarrow\hat{H}^{(k)}$ and, since the phases are arbitrary and independent of the interaction term, $\tilde{H}^{\prime(k)}(t)\rightarrow\hat{H}^{(k)}+(-1)^k\hbar\sum_{j=1}^{d^{(k)}}\left(\frac{d\theta_{j}(t)}{dt}\right)|\bar{\varphi}_{j}^{(k)}(t)\rangle\langle\bar{\varphi}_{j}^{(k)}(t)|$. Notice that, while the former is the time-translation generator of the basis $\{|\bar{\varphi}_{j}^{(k)}(t)\rangle\}_{j}$, the latter is the one relative to the basis $\{e^{(-1)^{k-1}i\theta_{j}(t)}|\bar{\varphi}_{j}^{(k)}(t)\rangle\}_{j}$. More importantly, it is clear that, in general, not all local effective Hamiltonians (phases) are suitable for correctly describing local energy measurements in the limit behaviour of non-interacting systems. Nevertheless, this is achieved iff $\frac{d}{dt}\theta_{j}(t)=\alpha$ for all $j$, since additive constants only shift the bare Hamiltonian spectrum. Thus, as long all the phases are equal linear functions of time, we have a physically consistent set of gauges satisfying $\tilde{H}^{\prime(k)}(t)=\tilde{H}^{(k)}(t)+(-1)^k\hbar \alpha\hat{1}^{(k)}$ and $\langle\tilde{H}^{\prime(k)}(t)\rangle=\langle\tilde{H}^{(k)}(t)\rangle+(-1)^k\hbar \alpha$ in a way that the remaining freedom is also guaranteed to maintain the spectral gaps of the local effective Hamiltonians invariant. Despite being local operators, observe that such a procedure requires knowledge of the ``whole'' in order to determine the ``parts'' energetics, i.e., obtaining the physical local effective Hamiltonians demands knowledge about the interaction of the parts (a non-local property) to fix the correct physical phase gauge. Hence, in order to construct a local and consistent energy description for the parts, one cannot rely solely on local features (see Reference \cite{rodrigo} for a recent discussion).


In summary, by restricting ourselves to gauges such that $\left\{ \frac{d}{dt}\theta_{j}(t)=\alpha\right\} _{j}$, where $\alpha\in\mathbb{R}$, one can identify the expectation values $\langle\tilde{H}^{(k)}(t)\rangle$ as the local physical internal energies up to an additive constant, $U^{(k)}(t):=\langle\tilde{H}^{(k)}(t)\rangle$ and $U^{\prime(k)}(t)=U^{(k)}(t)+(-1)^k\hbar \alpha$. In this sense, even though different gauges provide distinct absolute internal energy values, energy measurements differences remain identical.


\section{Conclusion}

In this work, we introduced a novel formalism suitable for describing the energetics within arbitrary autonomous quantum systems. Despite being presented for pure quantum states, the generalization for mixed ones is straightforward. The formal procedure is based on the well-known Schmidt decomposition given by equation (\ref{SchmidtDecomposition}), which provides the basis for the identification of the local effective Hamiltonians defined by equation (\ref{LocalEffectiveHamiltonian}). We highlighted the fact that these operators possess the desired properties for being considered suitable candidates for characterizing the subsystem's internal energies, i.e., their expectation values are local quantities and naturally satisfy the classical thermodynamic notion of energy additivity, as shown in equation (\ref{energy balance}). In contrast with currently used approaches, such a framework is exact, treats both partitions on equal footing and does not rely on approximations and additional restrictive hypotheses, i.e., particular coupling regimes, convenient Hamiltonian structures, semi-classical description and specific types of dynamics. 

Along with such an identification, we verified the existence of  an internal phase gauge freedom corresponding to the phases choices $\{\theta_{j}(t)\}_{j}$ within the Schmidt decomposition procedure. As shown in equations (\ref{Local effective Hamiltonian gauge change}, \ref{Expectation values Local effective Hamiltonian gauge change}), such arbitrary choice affects the structure of the local effective Hamiltonians and, in general, manifests as a time-dependent contribution to their expectation values. Nevertheless, despite the mathematical freedom, physical consistency during gauges transformations is only achieved for phases such that $\frac{d}{dt}\theta_{j}(t)=\alpha$ for all $j$, where $\alpha\in\mathbb{R}$. In this case, we have $\tilde{H}^{\prime(k)}(t)=\tilde{H}^{(k)}(t)+(-1)^k\hbar \alpha\hat{1}^{(k)}$ and $\langle\tilde{H}^{\prime(k)}(t)\rangle=\langle\tilde{H}^{(k)}(t)\rangle+(-1)^k\hbar \alpha$, and the limit behaviour $\tilde{H}^{\prime(k)}(t)\rightarrow\hat{H}^{(k)}+(-1)^k\hbar\alpha\hat{1}^{(k)}$ is satisfied for $\hat{H}_{int}\rightarrow0$.  Thus, by identifying the local internal energies as $U^{(k)}(t):=\langle\tilde{H}^{(k)}(t)\rangle$, different gauges would provide distinct absolute values for these quantities but equivalent spectral gaps. In this sense, the energy shift $(-1)^k\hbar \alpha$ is analogous to the classical thermodynamic freedom in the definition of internal energy \cite{Prigogine}.

In the current status of quantum thermodynamics, where there is no general methodology suitable for describing the energetic exchanges of arbitrary interacting quantum systems on equal footing, the procedure devised here represents a viable alternative to fill this important gap and a promising route to characterize general quantum thermodynamic processes. In particular, providing means to describe scenarios that do not fall under the standard description regimes, especially for strongly coupled systems, and in the absence of external classical agents.

\section{Methods}

\paragraph{Local effective Hamiltonian}

Given the following spectral decomposition for the local bare Hamiltonian 
\begin{equation}
    \hat{H}^{(k)}\equiv\sum_{j=1}^{d^{(k)}}b_{j}^{(k)}|b_{j}^{(k)}\rangle\langle b_{j}^{(k)}|,
\end{equation}
where $\{b_{j}^{(k)}\}_{j}$ and $\{|b_{j}^{(k)}\rangle\}_{j}$ are its respective bare eigenenergies and eigenbasis. One can define the projections $\langle b_{j}^{(k)}|\varphi_{l}^{(k)}(t)\rangle:=r_{jl}^{(k)}(t)e^{-\frac{i}{\hbar}b_{j}^{(k)}t}$, such that
\begin{multline*}
  \langle b_{j}^{(k)}|\frac{d}{dt}|\varphi_{l}^{(k)}(t)\rangle=\left(\frac{d}{dt}r_{jl}^{(k)}(t)\right)e^{-\frac{i}{\hbar}b_{j}^{(k)}t} \\ -\frac{i}{\hbar}b_{j}^{(k)}r_{jl}^{(k)}(t)e^{-\frac{i}{\hbar}b_{j}^{(k)}t}. 
\end{multline*}
Also, given the orthonormality $\langle b_{\alpha}^{(k)}|b_{\beta}^{(k)}\rangle=\delta_{\alpha\beta}$, it is easy to see that
\begin{equation*}
    \sum_{l=1}^{d^{(k)}}r_{\alpha l}^{(k)}(t)\left(r_{\beta l}^{(k)}(t)\right)^{*}e^{\frac{i}{\hbar}\left(b_{\beta}^{(k)}-b_{\alpha}^{(k)}\right)t}=\delta_{\alpha\beta}.
\end{equation*}
Finally, by casting $\tilde{H}^{(k)}(t)$ in the bare eigenbasis representation and using the previous relations, one can rewrite the local effective Hamiltonian according to equation (\ref{LocalEffectiveHamiltonian2}), i.e.,
\begin{equation*}
    \tilde{H}^{(k)}(t)=\hat{H}^{(k)}+\hat{H}_{LS}^{(k)}(t)+\hat{H}_{X}^{(k)}(t),
\end{equation*}
where
\begin{equation*}
    \hat{H}_{LS}^{(k)}(t) \equiv i\hbar\sum_{j=1}^{d^{(k)}}\left(\sum_{l=1}^{d^{(k)}}\left(\frac{d}{dt}r_{jl}^{(k)}(t)\right)r_{jl}^{(k)*}(t)\right)|b_{j}^{(k)}\rangle\langle b_{j}^{(k)}|,\label{Hamiltoniano LS}
\end{equation*}
and
\begin{multline*}
  \hat{H}_{X}^{(k)}(t) \equiv i\hbar\sum_{j=1}^{d^{(k)}}\sum_{m\neq j}^{d^{(k)}}\left(\sum_{l=1}^{d^{(k)}}\frac{d}{dt}r_{jl}^{(k)}(t)r_{ml}^{(k)*}(t)\right)\times\\e^{\frac{i}{\hbar}\left(b_{m}^{(k)}-b_{j}^{(k)}\right)t}|b_{j}^{(k)}\rangle\langle b_{m}^{(k)}|.\label{Hamiltoniano X}
\end{multline*}


\paragraph{Local internal energies additivity}

Given equation (\ref{SchmidtDecomposition}) and (\ref{Schmidt basis dynamics}) for the Schmidt decomposition and the Schmidt basis dynamics, we have the following equation
\begin{multline*}
  i\hbar\frac{d}{dt}|\Psi(t)\rangle=\left(\tilde{H}^{(1)}(t)+\tilde{H}^{(2)}(t)\right)|\Psi(t)\rangle\\+\sum_{j=1}^{d^{(1)}}\left(i\hbar\frac{d}{dt}\lambda_{j}(t)\right)|\varphi_{j}^{(1)}(t)\rangle\otimes|\varphi_{j}^{(2)}(t)\rangle.
\end{multline*}
Thus, since $i\hbar\frac{d}{dt}|\Psi(t)\rangle=\hat{H}^{(0)}|\Psi(t)\rangle$, we obtain
\begin{multline*}
  \langle\Psi(t)|\hat{H}^{(0)}|\Psi(t)\rangle=\langle\Psi(t)|\left(\tilde{H}^{(1)}(t)+\tilde{H}^{(2)}(t)\right)|\Psi(t)\rangle\\+\langle\Psi(t)|\sum_{j=1}^{d^{(1)}}\left(i\hbar\frac{d}{dt}\lambda_{j}(t)\right)|\varphi_{j}^{(1)}(t)\rangle\otimes|\varphi_{j}^{(2)}(t)\rangle.
\end{multline*}
However, notice that due to normalization of $|\Psi(t)\rangle$, the second contribution is necessarily null. Therefore,
\begin{equation}
    \langle\hat{H}^{(0)}\rangle=\langle\tilde{H}^{(1)}(t)\rangle+\langle\tilde{H}^{(2)}(t)\rangle=U^{(0)}.
\end{equation}

\section{Acknowledgements}

AM and FB acknowledge financial support in part by the Coordenação de Aperfeiçoamento de Pessoal de Nível Superior - Brasil (CAPES) - Finance Code 001. FB is also supported by the Brazilian National Institute for Science and Technology of Quantum Information (INCT- IQ). The authors would like to thank Prof. Shang-Yung Wang for the fruitful discussions.

\bibliography{refs}

\providecommand{\noopsort}[1]{}\providecommand{\singleletter}[1]{#1}%
\begin{thebibliography}{38}%
\makeatletter
\providecommand \@ifxundefined [1]{%
 \@ifx{#1\undefined}
}%
\providecommand \@ifnum [1]{%
 \ifnum #1\expandafter \@firstoftwo
 \else \expandafter \@secondoftwo
 \fi
}%
\providecommand \@ifx [1]{%
 \ifx #1\expandafter \@firstoftwo
 \else \expandafter \@secondoftwo
 \fi
}%
\providecommand \natexlab [1]{#1}%
\providecommand \enquote  [1]{``#1''}%
\providecommand \bibnamefont  [1]{#1}%
\providecommand \bibfnamefont [1]{#1}%
\providecommand \citenamefont [1]{#1}%
\providecommand \href@noop [0]{\@secondoftwo}%
\providecommand \href [0]{\begingroup \@sanitize@url \@href}%
\providecommand \@href[1]{\@@startlink{#1}\@@href}%
\providecommand \@@href[1]{\endgroup#1\@@endlink}%
\providecommand \@sanitize@url [0]{\catcode `\\12\catcode `\$12\catcode
  `\&12\catcode `\#12\catcode `\^12\catcode `\_12\catcode `\%12\relax}%
\providecommand \@@startlink[1]{}%
\providecommand \@@endlink[0]{}%
\providecommand \url  [0]{\begingroup\@sanitize@url \@url }%
\providecommand \@url [1]{\endgroup\@href {#1}{\urlprefix }}%
\providecommand \urlprefix  [0]{URL }%
\providecommand \Eprint [0]{\href }%
\providecommand \doibase [0]{https://doi.org/}%
\providecommand \selectlanguage [0]{\@gobble}%
\providecommand \bibinfo  [0]{\@secondoftwo}%
\providecommand \bibfield  [0]{\@secondoftwo}%
\providecommand \translation [1]{[#1]}%
\providecommand \BibitemOpen [0]{}%
\providecommand \bibitemStop [0]{}%
\providecommand \bibitemNoStop [0]{.\EOS\space}%
\providecommand \EOS [0]{\spacefactor3000\relax}%
\providecommand \BibitemShut  [1]{\csname bibitem#1\endcsname}%
\let\auto@bib@innerbib\@empty
\bibitem [{\citenamefont {Jarzynski}(1997)}]{jarzynski1997nonequilibrium}%
  \BibitemOpen
  \bibfield  {author} {\bibinfo {author} {\bibfnamefont {C.}~\bibnamefont
  {Jarzynski}},\ }\bibfield  {title} {\bibinfo {title} {Nonequilibrium equality
  for free energy differences},\ }\href@noop {} {\bibfield  {journal} {\bibinfo
   {journal} {Physical Review Letters}\ }\textbf {\bibinfo {volume} {78}},\
  \bibinfo {pages} {2690} (\bibinfo {year} {1997})},\ \bibinfo {note}
  {\doi{10.1103/PhysRevLett.78.2690}}\BibitemShut {NoStop}%
\bibitem [{\citenamefont {Seifert}(2012)}]{seifert2012stochastic}%
  \BibitemOpen
  \bibfield  {author} {\bibinfo {author} {\bibfnamefont {U.}~\bibnamefont
  {Seifert}},\ }\bibfield  {title} {\bibinfo {title} {Stochastic
  thermodynamics, fluctuation theorems and molecular machines},\ }\href@noop {}
  {\bibfield  {journal} {\bibinfo  {journal} {Reports on Progress in Physics}\
  }\textbf {\bibinfo {volume} {75}},\ \bibinfo {pages} {126001} (\bibinfo
  {year} {2012})},\ \bibinfo {note}
  {\doi{10.1088/0034-4885/75/12/126001}}\BibitemShut {NoStop}%
\bibitem [{\citenamefont {Gemmer}\ \emph {et~al.}(2009)\citenamefont {Gemmer},
  \citenamefont {Michel},\ and\ \citenamefont {Mahler}}]{gemmer2009quantum}%
  \BibitemOpen
  \bibfield  {author} {\bibinfo {author} {\bibfnamefont {J.}~\bibnamefont
  {Gemmer}}, \bibinfo {author} {\bibfnamefont {M.}~\bibnamefont {Michel}},\
  and\ \bibinfo {author} {\bibfnamefont {G.}~\bibnamefont {Mahler}},\
  }\href@noop {} {\emph {\bibinfo {title} {Quantum thermodynamics}}},\ Lecture
  notes in physics, v. 784\ (\bibinfo  {publisher} {Springer},\ \bibinfo
  {address} {Cham},\ \bibinfo {year} {2009})\ \bibinfo {note}
  {\doi{10.1007/978-3-540-70510-9}}\BibitemShut {NoStop}%
\bibitem [{\citenamefont {Vinjanampathy}\ and\ \citenamefont
  {Anders}(2016)}]{vinjanampathy2016quantum}%
  \BibitemOpen
  \bibfield  {author} {\bibinfo {author} {\bibfnamefont {S.}~\bibnamefont
  {Vinjanampathy}}\ and\ \bibinfo {author} {\bibfnamefont {J.}~\bibnamefont
  {Anders}},\ }\bibfield  {title} {\bibinfo {title} {Quantum thermodynamics},\
  }\href@noop {} {\bibfield  {journal} {\bibinfo  {journal} {Contemporary
  Physics}\ }\textbf {\bibinfo {volume} {57}},\ \bibinfo {pages} {545}
  (\bibinfo {year} {2016})},\ \bibinfo {note}
  {\doi{10.1080/00107514.2016.1201896}}\BibitemShut {NoStop}%
\bibitem [{\citenamefont {Binder}\ \emph {et~al.}(2018)\citenamefont {Binder},
  \citenamefont {Correa}, \citenamefont {Gogolin}, \citenamefont {Anders},\
  and\ \citenamefont {Adesso}}]{binder2018thermodynamics}%
  \BibitemOpen
  \bibfield  {author} {\bibinfo {author} {\bibfnamefont {F.}~\bibnamefont
  {Binder}}, \bibinfo {author} {\bibfnamefont {L.~A.}\ \bibnamefont {Correa}},
  \bibinfo {author} {\bibfnamefont {C.}~\bibnamefont {Gogolin}}, \bibinfo
  {author} {\bibfnamefont {J.}~\bibnamefont {Anders}},\ and\ \bibinfo {author}
  {\bibfnamefont {G.}~\bibnamefont {Adesso}},\ }\href@noop {} {\emph {\bibinfo
  {title} {Thermodynamics in the quantum regime}}},\ Fundamental theories of
  physics, v. 195\ (\bibinfo  {publisher} {Springer},\ \bibinfo {address}
  {Cham},\ \bibinfo {year} {2018})\ \bibinfo {note}
  {\doi{10.1007/978-3-319-99046-0}}\BibitemShut {NoStop}%
\bibitem [{\citenamefont {Deffner}\ and\ \citenamefont
  {Campbell}(2019)}]{deffner2019quantum}%
  \BibitemOpen
  \bibfield  {author} {\bibinfo {author} {\bibfnamefont {S.}~\bibnamefont
  {Deffner}}\ and\ \bibinfo {author} {\bibfnamefont {S.}~\bibnamefont
  {Campbell}},\ }\href@noop {} {\emph {\bibinfo {title} {Quantum
  thermodynamics}}}\ (\bibinfo  {publisher} {Morgan $\&$ Claypool Publishers},\
  \bibinfo {address} {San Rafael, CA},\ \bibinfo {year} {2019})\ \bibinfo
  {note} {\doi{10.1088/2053-2571/ab21c6}}\BibitemShut {NoStop}%
\bibitem [{\citenamefont {Batalh{\~a}o}\ \emph {et~al.}(2014)\citenamefont
  {Batalh{\~a}o}, \citenamefont {Souza}, \citenamefont {Mazzola}, \citenamefont
  {Auccaise}, \citenamefont {Sarthour}, \citenamefont {Oliveira}, \citenamefont
  {Goold}, \citenamefont {De~Chiara}, \citenamefont {Paternostro},\ and\
  \citenamefont {Serra}}]{batalhao2014experimental}%
  \BibitemOpen
  \bibfield  {author} {\bibinfo {author} {\bibfnamefont {T.~B.}\ \bibnamefont
  {Batalh{\~a}o}}, \bibinfo {author} {\bibfnamefont {A.~M.}\ \bibnamefont
  {Souza}}, \bibinfo {author} {\bibfnamefont {L.}~\bibnamefont {Mazzola}},
  \bibinfo {author} {\bibfnamefont {R.}~\bibnamefont {Auccaise}}, \bibinfo
  {author} {\bibfnamefont {R.~S.}\ \bibnamefont {Sarthour}}, \bibinfo {author}
  {\bibfnamefont {I.~S.}\ \bibnamefont {Oliveira}}, \bibinfo {author}
  {\bibfnamefont {J.}~\bibnamefont {Goold}}, \bibinfo {author} {\bibfnamefont
  {G.}~\bibnamefont {De~Chiara}}, \bibinfo {author} {\bibfnamefont
  {M.}~\bibnamefont {Paternostro}},\ and\ \bibinfo {author} {\bibfnamefont
  {R.~M.}\ \bibnamefont {Serra}},\ }\bibfield  {title} {\bibinfo {title}
  {Experimental reconstruction of work distribution and study of fluctuation
  relations in a closed quantum system},\ }\href@noop {} {\bibfield  {journal}
  {\bibinfo  {journal} {Physical Review Letters}\ }\textbf {\bibinfo {volume}
  {113}},\ \bibinfo {pages} {140601} (\bibinfo {year} {2014})},\ \bibinfo
  {note} {\doi{10.1103/PhysRevLett.113.140601}}\BibitemShut {NoStop}%
\bibitem [{\citenamefont {An}\ \emph {et~al.}(2015)\citenamefont {An},
  \citenamefont {Zhang}, \citenamefont {Um}, \citenamefont {Lv}, \citenamefont
  {Lu}, \citenamefont {Zhang}, \citenamefont {Yin}, \citenamefont {Quan},\ and\
  \citenamefont {Kim}}]{an2015experimental}%
  \BibitemOpen
  \bibfield  {author} {\bibinfo {author} {\bibfnamefont {S.}~\bibnamefont
  {An}}, \bibinfo {author} {\bibfnamefont {J.-N.}\ \bibnamefont {Zhang}},
  \bibinfo {author} {\bibfnamefont {M.}~\bibnamefont {Um}}, \bibinfo {author}
  {\bibfnamefont {D.}~\bibnamefont {Lv}}, \bibinfo {author} {\bibfnamefont
  {Y.}~\bibnamefont {Lu}}, \bibinfo {author} {\bibfnamefont {J.}~\bibnamefont
  {Zhang}}, \bibinfo {author} {\bibfnamefont {Z.-Q.}\ \bibnamefont {Yin}},
  \bibinfo {author} {\bibfnamefont {H.}~\bibnamefont {Quan}},\ and\ \bibinfo
  {author} {\bibfnamefont {K.}~\bibnamefont {Kim}},\ }\bibfield  {title}
  {\bibinfo {title} {Experimental test of the quantum jarzynski equality with a
  trapped-ion system},\ }\href@noop {} {\bibfield  {journal} {\bibinfo
  {journal} {Nature Physics}\ }\textbf {\bibinfo {volume} {11}},\ \bibinfo
  {pages} {193} (\bibinfo {year} {2015})},\ \bibinfo {note}
  {\doi{10.1038/nphys3197}}\BibitemShut {NoStop}%
\bibitem [{\citenamefont {Ro{\ss}nagel}\ \emph {et~al.}(2016)\citenamefont
  {Ro{\ss}nagel}, \citenamefont {Dawkins}, \citenamefont {Tolazzi},
  \citenamefont {Abah}, \citenamefont {Lutz}, \citenamefont {Schmidt-Kaler},\
  and\ \citenamefont {Singer}}]{rossnagel2016single}%
  \BibitemOpen
  \bibfield  {author} {\bibinfo {author} {\bibfnamefont {J.}~\bibnamefont
  {Ro{\ss}nagel}}, \bibinfo {author} {\bibfnamefont {S.~T.}\ \bibnamefont
  {Dawkins}}, \bibinfo {author} {\bibfnamefont {K.~N.}\ \bibnamefont
  {Tolazzi}}, \bibinfo {author} {\bibfnamefont {O.}~\bibnamefont {Abah}},
  \bibinfo {author} {\bibfnamefont {E.}~\bibnamefont {Lutz}}, \bibinfo {author}
  {\bibfnamefont {F.}~\bibnamefont {Schmidt-Kaler}},\ and\ \bibinfo {author}
  {\bibfnamefont {K.}~\bibnamefont {Singer}},\ }\bibfield  {title} {\bibinfo
  {title} {A single-atom heat engine},\ }\href@noop {} {\bibfield  {journal}
  {\bibinfo  {journal} {Science}\ }\textbf {\bibinfo {volume} {352}},\ \bibinfo
  {pages} {325} (\bibinfo {year} {2016})},\ \bibinfo {note}
  {\doi{10.1126/science.aad6320}}\BibitemShut {NoStop}%
\bibitem [{\citenamefont {Klatzow}\ \emph {et~al.}(2019)\citenamefont
  {Klatzow}, \citenamefont {Becker}, \citenamefont {Ledingham}, \citenamefont
  {Weinzetl}, \citenamefont {Kaczmarek}, \citenamefont {Saunders},
  \citenamefont {Nunn}, \citenamefont {Walmsley}, \citenamefont {Uzdin},\ and\
  \citenamefont {Poem}}]{klatzow2019experimental}%
  \BibitemOpen
  \bibfield  {author} {\bibinfo {author} {\bibfnamefont {J.}~\bibnamefont
  {Klatzow}}, \bibinfo {author} {\bibfnamefont {J.~N.}\ \bibnamefont {Becker}},
  \bibinfo {author} {\bibfnamefont {P.~M.}\ \bibnamefont {Ledingham}}, \bibinfo
  {author} {\bibfnamefont {C.}~\bibnamefont {Weinzetl}}, \bibinfo {author}
  {\bibfnamefont {K.~T.}\ \bibnamefont {Kaczmarek}}, \bibinfo {author}
  {\bibfnamefont {D.~J.}\ \bibnamefont {Saunders}}, \bibinfo {author}
  {\bibfnamefont {J.}~\bibnamefont {Nunn}}, \bibinfo {author} {\bibfnamefont
  {I.~A.}\ \bibnamefont {Walmsley}}, \bibinfo {author} {\bibfnamefont
  {R.}~\bibnamefont {Uzdin}},\ and\ \bibinfo {author} {\bibfnamefont
  {E.}~\bibnamefont {Poem}},\ }\bibfield  {title} {\bibinfo {title}
  {Experimental demonstration of quantum effects in the operation of
  microscopic heat engines},\ }\href@noop {} {\bibfield  {journal} {\bibinfo
  {journal} {Physical Review Letters}\ }\textbf {\bibinfo {volume} {122}},\
  \bibinfo {pages} {110601} (\bibinfo {year} {2019})},\ \bibinfo {note}
  {\doi{10.1103/PhysRevLett.122.110601}}\BibitemShut {NoStop}%
\bibitem [{\citenamefont {Alicki}(1979)}]{Alicki1979}%
  \BibitemOpen
  \bibfield  {author} {\bibinfo {author} {\bibfnamefont {R.}~\bibnamefont
  {Alicki}},\ }\bibfield  {title} {\bibinfo {title} {The quantum open system as
  a model of the heat engine},\ }\href@noop {} {\bibfield  {journal} {\bibinfo
  {journal} {Journal of Physics A: \textnormal{mathematical and general}}\
  }\textbf {\bibinfo {volume} {12}},\ \bibinfo {pages} {L103} (\bibinfo {year}
  {1979})},\ \bibinfo {note} {\doi{10.1088/0305-4470/12/5/007}}\BibitemShut
  {NoStop}%
\bibitem [{\citenamefont {Weimer}\ \emph {et~al.}(2008)\citenamefont {Weimer},
  \citenamefont {Henrich}, \citenamefont {Rempp}, \citenamefont
  {Schr{\"o}der},\ and\ \citenamefont {Mahler}}]{Weimer2008}%
  \BibitemOpen
  \bibfield  {author} {\bibinfo {author} {\bibfnamefont {H.}~\bibnamefont
  {Weimer}}, \bibinfo {author} {\bibfnamefont {M.~J.}\ \bibnamefont {Henrich}},
  \bibinfo {author} {\bibfnamefont {F.}~\bibnamefont {Rempp}}, \bibinfo
  {author} {\bibfnamefont {H.}~\bibnamefont {Schr{\"o}der}},\ and\ \bibinfo
  {author} {\bibfnamefont {G.}~\bibnamefont {Mahler}},\ }\bibfield  {title}
  {\bibinfo {title} {Local effective dynamics of quantum systems: a generalized
  approach to work and heat},\ }\href@noop {} {\bibfield  {journal} {\bibinfo
  {journal} {EPL (Europhysics Letters)}\ }\textbf {\bibinfo {volume} {83}},\
  \bibinfo {pages} {30008} (\bibinfo {year} {2008})},\ \bibinfo {note}
  {\doi{10.1209/0295-5075/83/30008}}\BibitemShut {NoStop}%
\bibitem [{\citenamefont {Alipour}\ \emph {et~al.}(2016)\citenamefont
  {Alipour}, \citenamefont {Benatti}, \citenamefont {Bakhshinezhad},
  \citenamefont {Afsary}, \citenamefont {Marcantoni},\ and\ \citenamefont
  {Rezakhani}}]{Alipour2016a}%
  \BibitemOpen
  \bibfield  {author} {\bibinfo {author} {\bibfnamefont {S.}~\bibnamefont
  {Alipour}}, \bibinfo {author} {\bibfnamefont {F.}~\bibnamefont {Benatti}},
  \bibinfo {author} {\bibfnamefont {F.}~\bibnamefont {Bakhshinezhad}}, \bibinfo
  {author} {\bibfnamefont {M.}~\bibnamefont {Afsary}}, \bibinfo {author}
  {\bibfnamefont {S.}~\bibnamefont {Marcantoni}},\ and\ \bibinfo {author}
  {\bibfnamefont {A.~T.}\ \bibnamefont {Rezakhani}},\ }\bibfield  {title}
  {\bibinfo {title} {Correlations in quantum thermodynamics: heat, work, and
  entropy production},\ }\href@noop {} {\bibfield  {journal} {\bibinfo
  {journal} {Scientific Reports}\ }\textbf {\bibinfo {volume} {6}},\ \bibinfo
  {pages} {1} (\bibinfo {year} {2016})},\ \bibinfo {note}
  {\doi{10.1038/srep35568}}\BibitemShut {NoStop}%
\bibitem [{\citenamefont {Valente}\ \emph {et~al.}(2018)\citenamefont
  {Valente}, \citenamefont {Brito}, \citenamefont {Ferreira},\ and\
  \citenamefont {Werlang}}]{Valente2018}%
  \BibitemOpen
  \bibfield  {author} {\bibinfo {author} {\bibfnamefont {D.}~\bibnamefont
  {Valente}}, \bibinfo {author} {\bibfnamefont {F.}~\bibnamefont {Brito}},
  \bibinfo {author} {\bibfnamefont {R.}~\bibnamefont {Ferreira}},\ and\
  \bibinfo {author} {\bibfnamefont {T.}~\bibnamefont {Werlang}},\ }\bibfield
  {title} {\bibinfo {title} {Work on a quantum dipole by a single-photon
  pulse},\ }\href@noop {} {\bibfield  {journal} {\bibinfo  {journal} {Optics
  Letters}\ }\textbf {\bibinfo {volume} {43}},\ \bibinfo {pages} {2644}
  (\bibinfo {year} {2018})},\ \bibinfo {note}
  {\doi{10.1364/OL.43.002644}}\BibitemShut {NoStop}%
\bibitem [{\citenamefont {Colla}\ and\ \citenamefont
  {Breuer}(2022)}]{Colla2021a}%
  \BibitemOpen
  \bibfield  {author} {\bibinfo {author} {\bibfnamefont {A.}~\bibnamefont
  {Colla}}\ and\ \bibinfo {author} {\bibfnamefont {H.-P.}\ \bibnamefont
  {Breuer}},\ }\bibfield  {title} {\bibinfo {title} {Open-system approach to
  nonequilibrium quantum thermodynamics at arbitrary coupling},\ }\href@noop {}
  {\bibfield  {journal} {\bibinfo  {journal} {Physical Review A}\ }\textbf
  {\bibinfo {volume} {105}},\ \bibinfo {pages} {052216} (\bibinfo {year}
  {2022})},\ \bibinfo {note} {\doi{10.1103/PhysRevA.105.052216}}\BibitemShut
  {NoStop}%
\bibitem [{\citenamefont {Esposito}\ \emph {et~al.}(2010)\citenamefont
  {Esposito}, \citenamefont {Lindenberg},\ and\ \citenamefont {Van~den
  Broeck}}]{Esposito_2010}%
  \BibitemOpen
  \bibfield  {author} {\bibinfo {author} {\bibfnamefont {M.}~\bibnamefont
  {Esposito}}, \bibinfo {author} {\bibfnamefont {K.}~\bibnamefont
  {Lindenberg}},\ and\ \bibinfo {author} {\bibfnamefont {C.}~\bibnamefont
  {Van~den Broeck}},\ }\bibfield  {title} {\bibinfo {title} {Entropy production
  as correlation between system and reservoir},\ }\href@noop {} {\bibfield
  {journal} {\bibinfo  {journal} {New Journal of Physics}\ }\textbf {\bibinfo
  {volume} {12}},\ \bibinfo {pages} {013013} (\bibinfo {year} {2010})},\
  \bibinfo {note} {\doi{10.1088/1367-2630/12/1/013013}}\BibitemShut {NoStop}%
\bibitem [{\citenamefont {Polkovnikov}(2011)}]{Polkovnikov2011}%
  \BibitemOpen
  \bibfield  {author} {\bibinfo {author} {\bibfnamefont {A.}~\bibnamefont
  {Polkovnikov}},\ }\bibfield  {title} {\bibinfo {title} {Microscopic diagonal
  entropy and its connection to basic thermodynamic relations},\ }\href@noop {}
  {\bibfield  {journal} {\bibinfo  {journal} {Annals of Physics}\ }\textbf
  {\bibinfo {volume} {326}},\ \bibinfo {pages} {486} (\bibinfo {year}
  {2011})},\ \bibinfo {note} {\doi{10.1016/j.aop.2010.08.004}}\BibitemShut
  {NoStop}%
\bibitem [{\citenamefont {{\v{S}}afr{\'a}nek}\ \emph
  {et~al.}(2019)\citenamefont {{\v{S}}afr{\'a}nek}, \citenamefont {Deutsch},\
  and\ \citenamefont {Aguirre}}]{Entropy2019}%
  \BibitemOpen
  \bibfield  {author} {\bibinfo {author} {\bibfnamefont {D.}~\bibnamefont
  {{\v{S}}afr{\'a}nek}}, \bibinfo {author} {\bibfnamefont {J.~M.}\ \bibnamefont
  {Deutsch}},\ and\ \bibinfo {author} {\bibfnamefont {A.}~\bibnamefont
  {Aguirre}},\ }\bibfield  {title} {\bibinfo {title} {Quantum coarse-grained
  entropy and thermodynamics},\ }\href@noop {} {\bibfield  {journal} {\bibinfo
  {journal} {Physical Review A}\ }\textbf {\bibinfo {volume} {99}},\ \bibinfo
  {pages} {010101} (\bibinfo {year} {2019})},\ \bibinfo {note}
  {\doi{10.1103/PhysRevA.99.010101}}\BibitemShut {NoStop}%
\bibitem [{\citenamefont {Talkner}\ \emph {et~al.}(2007)\citenamefont
  {Talkner}, \citenamefont {Lutz},\ and\ \citenamefont
  {H{\"a}nggi}}]{Talkner2007}%
  \BibitemOpen
  \bibfield  {author} {\bibinfo {author} {\bibfnamefont {P.}~\bibnamefont
  {Talkner}}, \bibinfo {author} {\bibfnamefont {E.}~\bibnamefont {Lutz}},\ and\
  \bibinfo {author} {\bibfnamefont {P.}~\bibnamefont {H{\"a}nggi}},\ }\bibfield
   {title} {\bibinfo {title} {Fluctuation theorems: work is not an
  observable},\ }\href@noop {} {\bibfield  {journal} {\bibinfo  {journal}
  {Physical Review E}\ }\textbf {\bibinfo {volume} {75}},\ \bibinfo {pages}
  {050102} (\bibinfo {year} {2007})},\ \bibinfo {note}
  {\doi{10.1103/PhysRevE.75.050102}}\BibitemShut {NoStop}%
\bibitem [{\citenamefont {Elouard}\ \emph {et~al.}(2017)\citenamefont
  {Elouard}, \citenamefont {Herrera-Mart{\'\i}}, \citenamefont {Clusel},\ and\
  \citenamefont {Auffeves}}]{Elouard2016b}%
  \BibitemOpen
  \bibfield  {author} {\bibinfo {author} {\bibfnamefont {C.}~\bibnamefont
  {Elouard}}, \bibinfo {author} {\bibfnamefont {D.~A.}\ \bibnamefont
  {Herrera-Mart{\'\i}}}, \bibinfo {author} {\bibfnamefont {M.}~\bibnamefont
  {Clusel}},\ and\ \bibinfo {author} {\bibfnamefont {A.}~\bibnamefont
  {Auffeves}},\ }\bibfield  {title} {\bibinfo {title} {The role of quantum
  measurement in stochastic thermodynamics},\ }\href@noop {} {\bibfield
  {journal} {\bibinfo  {journal} {npj Quantum Information}\ }\textbf {\bibinfo
  {volume} {3}},\ \bibinfo {pages} {1} (\bibinfo {year} {2017})},\ \bibinfo
  {note} {\doi{10.1038/s41534-017-0008-4}}\BibitemShut {NoStop}%
\bibitem [{\citenamefont {Perarnau-Llobet}\ \emph {et~al.}(2017)\citenamefont
  {Perarnau-Llobet}, \citenamefont {B{\"a}umer}, \citenamefont {Hovhannisyan},
  \citenamefont {Huber},\ and\ \citenamefont {Acin}}]{Perarnau-Llobet2017}%
  \BibitemOpen
  \bibfield  {author} {\bibinfo {author} {\bibfnamefont {M.}~\bibnamefont
  {Perarnau-Llobet}}, \bibinfo {author} {\bibfnamefont {E.}~\bibnamefont
  {B{\"a}umer}}, \bibinfo {author} {\bibfnamefont {K.~V.}\ \bibnamefont
  {Hovhannisyan}}, \bibinfo {author} {\bibfnamefont {M.}~\bibnamefont
  {Huber}},\ and\ \bibinfo {author} {\bibfnamefont {A.}~\bibnamefont {Acin}},\
  }\bibfield  {title} {\bibinfo {title} {No-go theorem for the characterization
  of work fluctuations in coherent quantum systems},\ }\href@noop {} {\bibfield
   {journal} {\bibinfo  {journal} {Physical Review Letters}\ }\textbf {\bibinfo
  {volume} {118}},\ \bibinfo {pages} {070601} (\bibinfo {year} {2017})},\
  \bibinfo {note} {\doi{10.1103/PhysRevLett.118.070601}}\BibitemShut {NoStop}%
\bibitem [{\citenamefont {Ahmadi}\ \emph {et~al.}(2021)\citenamefont {Ahmadi},
  \citenamefont {Salimi},\ and\ \citenamefont {Khorashad}}]{Ahmadi2019a}%
  \BibitemOpen
  \bibfield  {author} {\bibinfo {author} {\bibfnamefont {B.}~\bibnamefont
  {Ahmadi}}, \bibinfo {author} {\bibfnamefont {S.}~\bibnamefont {Salimi}},\
  and\ \bibinfo {author} {\bibfnamefont {A.~S.}\ \bibnamefont {Khorashad}},\
  }\bibfield  {title} {\bibinfo {title} {Refined definitions of heat and work
  in quantum thermodynamics},\ }\href@noop {} {\bibfield  {journal} {\bibinfo
  {journal} {arXiv preprint arXiv:1912.01983}\ } (\bibinfo {year} {2021})},\
  \bibinfo {note} {\doi{10.48550/arXiv.1912.01983}}\BibitemShut {NoStop}%
\bibitem [{\citenamefont {Alipour}\ \emph {et~al.}(2022)\citenamefont
  {Alipour}, \citenamefont {Rezakhani}, \citenamefont {Chenu}, \citenamefont
  {del Campo},\ and\ \citenamefont {Ala-Nissila}}]{Alipour2019b}%
  \BibitemOpen
  \bibfield  {author} {\bibinfo {author} {\bibfnamefont {S.}~\bibnamefont
  {Alipour}}, \bibinfo {author} {\bibfnamefont {A.~T.}\ \bibnamefont
  {Rezakhani}}, \bibinfo {author} {\bibfnamefont {A.}~\bibnamefont {Chenu}},
  \bibinfo {author} {\bibfnamefont {A.}~\bibnamefont {del Campo}},\ and\
  \bibinfo {author} {\bibfnamefont {T.}~\bibnamefont {Ala-Nissila}},\
  }\bibfield  {title} {\bibinfo {title} {Entropy-based formulation of
  thermodynamics in arbitrary quantum evolution},\ }\href@noop {} {\bibfield
  {journal} {\bibinfo  {journal} {Physical Review A}\ }\textbf {\bibinfo
  {volume} {105}},\ \bibinfo {pages} {L040201} (\bibinfo {year} {2022})},\
  \bibinfo {note} {\doi{10.1103/PhysRevA.105.L040201}}\BibitemShut {NoStop}%
\bibitem [{\citenamefont {Lima~Bernardo}(2020)}]{Bernardo2020}%
  \BibitemOpen
  \bibfield  {author} {\bibinfo {author} {\bibfnamefont {B.}~\bibnamefont
  {Lima~Bernardo}},\ }\bibfield  {title} {\bibinfo {title} {Unraveling the role
  of coherence in the first law of quantum thermodynamics},\ }\href@noop {}
  {\bibfield  {journal} {\bibinfo  {journal} {Physical Review E}\ }\textbf
  {\bibinfo {volume} {102}},\ \bibinfo {pages} {062152} (\bibinfo {year}
  {2020})},\ \bibinfo {note} {\doi{10.1103/PhysRevE.102.062152}}\BibitemShut
  {NoStop}%
\bibitem [{\citenamefont {Kosloff}(2013)}]{kosloff2013quantum}%
  \BibitemOpen
  \bibfield  {author} {\bibinfo {author} {\bibfnamefont {R.}~\bibnamefont
  {Kosloff}},\ }\bibfield  {title} {\bibinfo {title} {Quantum thermodynamics: a
  dynamical viewpoint},\ }\href@noop {} {\bibfield  {journal} {\bibinfo
  {journal} {Entropy}\ }\textbf {\bibinfo {volume} {15}},\ \bibinfo {pages}
  {2100} (\bibinfo {year} {2013})},\ \bibinfo {note}
  {\doi{10.3390/e15062100}}\BibitemShut {NoStop}%
\bibitem [{\citenamefont {Kosloff}(2019)}]{kosloff2019quantum}%
  \BibitemOpen
  \bibfield  {author} {\bibinfo {author} {\bibfnamefont {R.}~\bibnamefont
  {Kosloff}},\ }\bibfield  {title} {\bibinfo {title} {Quantum thermodynamics
  and open-systems modeling},\ }\href@noop {} {\bibfield  {journal} {\bibinfo
  {journal} {Journal of chemical physics}\ }\textbf {\bibinfo {volume} {150}},\
  \bibinfo {pages} {204105} (\bibinfo {year} {2019})},\ \bibinfo {note}
  {\doi{10.1063/1.5096173}}\BibitemShut {NoStop}%
\bibitem [{\citenamefont {Dann}\ and\ \citenamefont
  {Kosloff}(2021{\natexlab{a}})}]{dann2021open}%
  \BibitemOpen
  \bibfield  {author} {\bibinfo {author} {\bibfnamefont {R.}~\bibnamefont
  {Dann}}\ and\ \bibinfo {author} {\bibfnamefont {R.}~\bibnamefont {Kosloff}},\
  }\bibfield  {title} {\bibinfo {title} {Open system dynamics from
  thermodynamic compatibility},\ }\href@noop {} {\bibfield  {journal} {\bibinfo
   {journal} {Physical Review Research}\ }\textbf {\bibinfo {volume} {3}},\
  \bibinfo {pages} {023006} (\bibinfo {year} {2021}{\natexlab{a}})},\ \bibinfo
  {note} {\doi{10.1103/PhysRevResearch.3.023006}}\BibitemShut {NoStop}%
\bibitem [{\citenamefont {Hossein-Nejad}\ \emph {et~al.}(2015)\citenamefont
  {Hossein-Nejad}, \citenamefont {O’Reilly},\ and\ \citenamefont
  {Olaya-Castro}}]{Hossein-Nejad2015}%
  \BibitemOpen
  \bibfield  {author} {\bibinfo {author} {\bibfnamefont {H.}~\bibnamefont
  {Hossein-Nejad}}, \bibinfo {author} {\bibfnamefont {E.~J.}\ \bibnamefont
  {O’Reilly}},\ and\ \bibinfo {author} {\bibfnamefont {A.}~\bibnamefont
  {Olaya-Castro}},\ }\bibfield  {title} {\bibinfo {title} {Work, heat and
  entropy production in bipartite quantum systems},\ }\href@noop {} {\bibfield
  {journal} {\bibinfo  {journal} {New Journal of Physics}\ }\textbf {\bibinfo
  {volume} {17}},\ \bibinfo {pages} {075014} (\bibinfo {year} {2015})},\
  \bibinfo {note} {\doi{10.1088/1367-2630/17/7/075014}}\BibitemShut {NoStop}%
\bibitem [{\citenamefont {Dann}\ and\ \citenamefont
  {Kosloff}(2021{\natexlab{b}})}]{dann2021quantum}%
  \BibitemOpen
  \bibfield  {author} {\bibinfo {author} {\bibfnamefont {R.}~\bibnamefont
  {Dann}}\ and\ \bibinfo {author} {\bibfnamefont {R.}~\bibnamefont {Kosloff}},\
  }\bibfield  {title} {\bibinfo {title} {Quantum thermo-dynamical construction
  for driven open quantum systems},\ }\href@noop {} {\bibfield  {journal}
  {\bibinfo  {journal} {Quantum}\ }\textbf {\bibinfo {volume} {5}},\ \bibinfo
  {pages} {590} (\bibinfo {year} {2021}{\natexlab{b}})},\ \bibinfo {note}
  {\doi{10.22331/q-2021-11-25-590}}\BibitemShut {NoStop}%
\bibitem [{\citenamefont {Esposito}\ \emph {et~al.}(2015)\citenamefont
  {Esposito}, \citenamefont {Ochoa},\ and\ \citenamefont
  {Galperin}}]{esposito2015nature}%
  \BibitemOpen
  \bibfield  {author} {\bibinfo {author} {\bibfnamefont {M.}~\bibnamefont
  {Esposito}}, \bibinfo {author} {\bibfnamefont {M.~A.}\ \bibnamefont
  {Ochoa}},\ and\ \bibinfo {author} {\bibfnamefont {M.}~\bibnamefont
  {Galperin}},\ }\bibfield  {title} {\bibinfo {title} {Nature of heat in
  strongly coupled open quantum systems},\ }\href@noop {} {\bibfield  {journal}
  {\bibinfo  {journal} {Physical Review B}\ }\textbf {\bibinfo {volume} {92}},\
  \bibinfo {pages} {235440} (\bibinfo {year} {2015})},\ \bibinfo {note}
  {\doi{10.1103/PhysRevB.92.235440}}\BibitemShut {NoStop}%
\bibitem [{\citenamefont {Dou}\ \emph {et~al.}(2018)\citenamefont {Dou},
  \citenamefont {Ochoa}, \citenamefont {Nitzan},\ and\ \citenamefont
  {Subotnik}}]{PhysRevB.98.134306}%
  \BibitemOpen
  \bibfield  {author} {\bibinfo {author} {\bibfnamefont {W.}~\bibnamefont
  {Dou}}, \bibinfo {author} {\bibfnamefont {M.~A.}\ \bibnamefont {Ochoa}},
  \bibinfo {author} {\bibfnamefont {A.}~\bibnamefont {Nitzan}},\ and\ \bibinfo
  {author} {\bibfnamefont {J.~E.}\ \bibnamefont {Subotnik}},\ }\bibfield
  {title} {\bibinfo {title} {Universal approach to quantum thermodynamics in
  the strong coupling regime},\ }\href@noop {} {\bibfield  {journal} {\bibinfo
  {journal} {Physical Review B}\ }\textbf {\bibinfo {volume} {98}},\ \bibinfo
  {pages} {134306} (\bibinfo {year} {2018})},\ \bibinfo {note}
  {\doi{10.1103/PhysRevB.98.134306}}\BibitemShut {NoStop}%
\bibitem [{\citenamefont {Perarnau-Llobet}\ \emph {et~al.}(2018)\citenamefont
  {Perarnau-Llobet}, \citenamefont {Wilming}, \citenamefont {Riera},
  \citenamefont {Gallego},\ and\ \citenamefont {Eisert}}]{perarnau2018strong}%
  \BibitemOpen
  \bibfield  {author} {\bibinfo {author} {\bibfnamefont {M.}~\bibnamefont
  {Perarnau-Llobet}}, \bibinfo {author} {\bibfnamefont {H.}~\bibnamefont
  {Wilming}}, \bibinfo {author} {\bibfnamefont {A.}~\bibnamefont {Riera}},
  \bibinfo {author} {\bibfnamefont {R.}~\bibnamefont {Gallego}},\ and\ \bibinfo
  {author} {\bibfnamefont {J.}~\bibnamefont {Eisert}},\ }\bibfield  {title}
  {\bibinfo {title} {Strong coupling corrections in quantum thermodynamics},\
  }\href@noop {} {\bibfield  {journal} {\bibinfo  {journal} {Physical Review
  Letters}\ }\textbf {\bibinfo {volume} {120}},\ \bibinfo {pages} {120602}
  (\bibinfo {year} {2018})},\ \bibinfo {note}
  {\doi{10.1103/PhysRevLett.120.120602}}\BibitemShut {NoStop}%
\bibitem [{\citenamefont {Strasberg}(2019)}]{strasberg2019repeated}%
  \BibitemOpen
  \bibfield  {author} {\bibinfo {author} {\bibfnamefont {P.}~\bibnamefont
  {Strasberg}},\ }\bibfield  {title} {\bibinfo {title} {Repeated interactions
  and quantum stochastic thermodynamics at strong coupling},\ }\href@noop {}
  {\bibfield  {journal} {\bibinfo  {journal} {Physical Review Letters}\
  }\textbf {\bibinfo {volume} {123}},\ \bibinfo {pages} {180604} (\bibinfo
  {year} {2019})},\ \bibinfo {note}
  {\doi{10.1103/PhysRevLett.123.180604}}\BibitemShut {NoStop}%
\bibitem [{\citenamefont {Rivas}(2020)}]{Rivas2020}%
  \BibitemOpen
  \bibfield  {author} {\bibinfo {author} {\bibfnamefont {{\'A}.}~\bibnamefont
  {Rivas}},\ }\bibfield  {title} {\bibinfo {title} {Strong coupling
  thermodynamics of open quantum systems},\ }\href@noop {} {\bibfield
  {journal} {\bibinfo  {journal} {Physical Review Letters}\ }\textbf {\bibinfo
  {volume} {124}},\ \bibinfo {pages} {160601} (\bibinfo {year} {2020})},\
  \bibinfo {note} {\doi{10.1103/PhysRevLett.124.160601}}\BibitemShut {NoStop}%
\bibitem [{\citenamefont {Nielsen}\ and\ \citenamefont
  {Chuang}(2010)}]{nielsen10}%
  \BibitemOpen
  \bibfield  {author} {\bibinfo {author} {\bibfnamefont {M.~A.}\ \bibnamefont
  {Nielsen}}\ and\ \bibinfo {author} {\bibfnamefont {I.~L.}\ \bibnamefont
  {Chuang}},\ }\href@noop {} {\emph {\bibinfo {title} {Quantum computation and
  quantum information}}}\ (\bibinfo  {publisher} {Cambridge University Press},\
  \bibinfo {address} {Cambridge, UK},\ \bibinfo {year} {2010})\BibitemShut
  {NoStop}%
\bibitem [{Note1()}]{Note1}%
  \BibitemOpen
  \bibinfo {note} {In a private communication, Prof. Shang-Yung Wang presented
  an approach based on the introduction of gauge potentials such that the local
  effective Hamiltonian is gauge invariant. Even though this feature is quite
  neat, it does not preserve the association of the Hamiltonian as the
  time-translation generator, differing from the approach adopted
  here.}\BibitemShut {Stop}%
\bibitem [{\citenamefont {Neves}\ and\ \citenamefont {Brito}(2022)}]{rodrigo}%
  \BibitemOpen
  \bibfield  {author} {\bibinfo {author} {\bibfnamefont {L.~R.~T.}\
  \bibnamefont {Neves}}\ and\ \bibinfo {author} {\bibfnamefont
  {F.}~\bibnamefont {Brito}},\ }\bibfield  {title} {\bibinfo {title} {A
  constraint on local definitions of quantum internal energy},\ }\href@noop {}
  {\bibfield  {journal} {\bibinfo  {journal} {arXiv preprint arXiv:2205.04457}\
  } (\bibinfo {year} {2022})},\ \bibinfo {note}
  {\doi{10.48550/arXiv.2205.04457}}\BibitemShut {NoStop}%
\bibitem [{\citenamefont {Kondepudi}\ and\ \citenamefont
  {Prigogine}(2015)}]{Prigogine}%
  \BibitemOpen
  \bibfield  {author} {\bibinfo {author} {\bibfnamefont {D.}~\bibnamefont
  {Kondepudi}}\ and\ \bibinfo {author} {\bibfnamefont {I.}~\bibnamefont
  {Prigogine}},\ }\href@noop {} {\emph {\bibinfo {title} {Modern
  thermodynamics}}},\ \bibinfo {edition} {2nd}\ ed.\ (\bibinfo  {publisher}
  {John Wiley},\ \bibinfo {address} {New York},\ \bibinfo {year}
  {2015})\BibitemShut {NoStop}%
\end{thebibliography}%

\end{document}